\documentclass[11pt,a4paper]{article}

\usepackage[margin=1in]{geometry}
\usepackage{setspace}
\usepackage{authblk}
\usepackage{graphicx}
\usepackage{amsmath,amssymb}
\usepackage{bm}
\usepackage{siunitx}
\usepackage{booktabs}
\usepackage{subcaption}
\usepackage{hyperref}
\usepackage{authblk}

\usepackage{biblatex}
\begin{document}

\title{\textbf{Deep learning water-unsuppressed MRSI at ultra-high field for simultaneous quantitative metabolic, susceptibility and myelin water imaging}}

\author[1,2,3]{Paul J. Weiser}
\author[4]{Jiye Kim}
\author[4]{Jongho Lee}
\author[5,6]{Amirmohammad Shamaei}
\author[1,2]{Gulnur Ungan}
\author[1,2]{Malte Hoffmann}
\author[7]{Antoine Klauser}
\author[1,2]{Berkin Bilgic}
\author[1,2]{Ovidiu C. Andronesi}

\affil[1]{Athinoula A. Martinos Center for Biomedical Imaging, Massachusetts General Hospital, Boston, MA, USA}
\affil[2]{Department of Radiology, Massachusetts General Hospital, Harvard Medical School, Boston, MA, USA}
\affil[3]{Computational Imaging Research Lab, Department of Biomedical Imaging and Image-guided Therapy, Medical University of Vienna, Vienna, Austria}
\affil[4]{Department of Electrical and Computer Engineering, Seoul National University, Seoul, Republic of Korea}
\affil[5]{Electrical and Software Engineering, University of Calgary, Calgary, Canada}
\affil[6]{Hotchkiss Brain Institute, University of Calgary, Calgary, Canada}
\affil[7]{Advanced Clinical Imaging Technology, Siemens Healthineers International AG, Lausanne, Switzerland}

\date{} 

\maketitle

\begin{abstract}

\textbf{Purpose} \\
Magnetic Resonance Spectroscopic Imaging (MRSI) maps endogenous brain metabolism while suppressing the overwhelming water signal.
Water-unsuppressed MRSI (wu-MRSI) allows simultaneous imaging of water and metabolites,
but large water sidebands cause challenges for metabolic fitting.
We developed an end-to-end deep-learning pipeline to overcome these challenges at ultra-high field.

\medskip
\textbf{Methods} \\
Fast high-resolution wu-MRSI was acquired at 7T with non-cartesian ECCENTRIC sampling and ultra-short echo time.
A water and lipid removal network (WALINET+) was developed to remove lipids, water signal, and sidebands.
MRSI reconstruction was performed by DeepER and a physics-informed network for metabolite fitting.
Water signal was used for absolute metabolite quantification, quantitative susceptibility mapping (QSM),
and myelin water fraction imaging (MWF).

\medskip
\textbf{Results} \\
WALINET+ provided the lowest NRMSE ($<2\%$) in simulations and in vivo the smallest bias ($<20\%$)
and limits-of-agreement ($\pm63\%$) between wu-MRSI and ws-MRSI scans.
Several metabolites such as creatine and glutamate showed higher SNR in wu-MRSI.
QSM and MWF obtained from wu-MRSI and GRE showed good agreement with
$0\,\mathrm{ppm}/5.5\%$ bias and $\pm0.05\,\mathrm{ppm}/\pm12.75\%$ limits-of-agreement.

\medskip
\textbf{Conclusion} \\
High-quality metabolic, QSM, and MWF mapping of the human brain can be obtained simultaneously
by ECCENTRIC wu-MRSI at 7T with 2\,mm isotropic resolution in 12\,min.
WALINET+ robustly removes water sidebands while preserving metabolite signal,
eliminating the need for water suppression and separate water acquisitions.

\end{abstract}

\section*{Keywords}
wu-MRSI; QSM; MWF; WALINET+; DeepER; ultra-high field

\maketitle




\section{Introduction}\label{sec1}
${}^1$H-MRSI is a label-free non-invasive molecular imaging method that provides rich information about the intrinsic neurochemistry of the brain. This has great specificity to investigate healthy and pathological conditions\cite{oz2014clinical}, such as brain tumors \cite{andronesi2018pharmacodynamics} and neuropsychiatric diseases including amyotrophic lateral sclerosis\cite{andronesi2020imaging}, multiple sclerosis\cite{heckova2022ms}, Alzheimer's Disease\cite{hu2024ad}. In the brain, the concentration of metabolites (1-20 mM) detectable by MRSI is 3-4 orders of magnitude lower than the water concentration (30-55 M). Due to this large dynamic range between metabolites and water, MRSI data acquisition is typically performed using water suppression techniques such as WET or VAPOR \cite{ogg1994wet, tkavc1999vivo, tkavc2001vivo, tkac2005methodology, tkavc2021water}. However, these techniques increase the minimum repetition time and SAR, prolonging the total scan time. In addition, water suppression is sensitive to $B_0$ and $B_1+$ inhomogeneity, resulting in variable water suppression efficiency across the brain, and can decrease the metabolite signal either by direct suppression through the RF saturation or by chemical exchange of labile protons between metabolites and water. On the other hand, suppression of the water signal discards useful information. The water signal is needed for absolute metabolite quantification and can be used to probe many tissue properties relevant to biological investigations. Hence, separate acquisitions are required for the water signal, which prolongs the total scan time and is susceptible to misalignment between water and metabolite data due to head motion.   

Water-unsuppressed MRSI (wu-MRSI) \cite{dong2015proton} provides an efficient way to simultaneously acquire both the metabolites and the water signals, but requires solving the difficult technical challenge posed by the large water sidebands that overlap metabolite spectral peaks. To deal with the water signal in wu-MRSI, both data acquisition and data processing approaches have been explored \cite{barkhuijsen1987improved, serrai2002localized, le2014fid, dreher2005new, chadzynski2010chemical}. Data acquisition using metabolite cycling \cite{dreher2005new, chang2018non} relies on two separate acquisitions where the metabolite signal is selectively inverted in one of the acquisitions and the water signal is removed by subtraction of the two acquisitions. However, metabolite cycling doubles the acquisition time and increases SAR and minimum repetition time, further prolonging the total scan time. In addition, it is susceptible to $B_0$ and $B_1+$ inhomogeneity and motion, resulting in imperfect water signal cancellation and less averaging of the metabolite signal. 

Data processing methods that remove the main water signal and its sidebands can achieve more efficient wu-MRSI than metabolite cycling. Efficient processing methods have been shown for nuisance signal removal for ws-MRSI\cite{barkhuijsen1987improved,ma2016removal,nagaraja2018tensor,lin2019water,shamaei2024water}, however for water sideband removal in wu-MRSI there are significant challenges and only a few studies have investigated this avenue \cite{peng2018simultaneous,guo2021simultaneous}, with so far no established method for widespread adoption. In particular, removing the water sidebands while reliably preserving metabolite signals is more difficult than removing the main water peak for which established methods such as HLSVD exist \cite{barkhuijsen1987improved}. Water sidebands are induced by gradient switching which cause mechanical vibrations of the gradient coil that modulate the main magnetic field $B_0$. This results in water sidebands at several frequencies across the spectral bandwidth. The water sidebands have amplitudes of 1-2\% of the main water peak, but are larger than the metabolite peaks. The amplitude of the water sidebands increases with larger gradient amplitude and shorter echo times, rendering metabolite imaging difficult for short-echo sequences. 

To date, most processing methods for water sideband removal use either simple subtraction \cite{chadzynski2010chemical} or physical modeling \cite{nixon2008compensation} as damped oscillations. The latter approach requires detailed knowledge of vibration frequencies, amplitudes and their exponential decay time constants; determining those parameters is a tedious task, and they can vary between scanners. Furthermore, often the physical modeling does not capture the full diversity of sidebands in data acquisition because of incomplete knowledge of the vibration modes, and hence results in incomplete water sideband removal.

More advanced processing methods \cite{peng2018simultaneous,guo2021simultaneous} split the time domain signal in two regions: i) the initial time points (short-echo region) which is used to estimate the water sidebands based on the assumption that the decay of the sideband signal is faster than the metabolite signal, and ii) the later time points (long-echo region) used for metabolite signal. Such an approach may be sub-optimal for very short echo ws-MRSI since discarding the initial time points lowers metabolite SNR and results in a linear phase over the chemical shift range that complicates the spectral pattern and fitting. Although back-prediction \cite{koehl1999linear} can recover some of the initial time points as employed in ex-vivo NMR spectroscopy, the lower data quality of in-vivo MRS limits the accuracy of the prediction, especially when a large number of time points need to be predicted.  

The problem of water sideband removal from wu-MRSI is well suited to a deep learning approach that can learn the comprehensive distribution of sidebands from a large amount of wu-MRSI data acquired with a variety of imaging protocols. For this purpose we extended the performance of the WALINET network \cite{weiser2025walinet}, which we previously developed for removal of lipid and residual water signals from water-suppressed MRSI. WALINET+ extends WALINET to remove water sidebands in addition to the main water peak and lipids from very short-echo water-unsuppressed MRSI data. WALINET+ performance differs from previous approaches\cite{barkhuijsen1987improved,nixon2008compensation,peng2018simultaneous,guo2021simultaneous} as follows: it 1) does not require laborious mapping of the vibration modes for physical modeling, 2) uses all the time-points for metabolite processing, 3) can be easily adapted to a change in scanner hardware and pulse sequence, 4) integrates the removal of the main water signal, the water sidebands and the lipid signals into a single processing step. In particular, the latter aspect significantly simplifies the processing pipeline compared to prior approaches \cite{barkhuijsen1987improved,chadzynski2010chemical,bilgic2013lipid} that rely on three separate steps. In addition to the processing pipeline, we optimized the gradient waveforms in the pulse sequence to reduce mechanical vibrations and the resulting water sidebands.    

The readily available water signal from the water-unsuppressed MRSI can be used for several important applications: 1) as an internal reference for absolute metabolite quantification, 2) quantitative susceptibility mapping (QSM) \cite{peng2018simultaneous}, and 3) myelin water fraction (MWF) imaging \cite{guo2021simultaneous}. In particular, wu-MRSI can provide hundreds of echo times with sub-millisecond increments that can probe T2* decay in great detail which can be used for accurate modeling of local-field susceptibility and water pool relaxation for QSM and MWF imaging, respectively. Multiple compartment relaxometry can probe the water pool bound to myelin, which is a sensitive imaging biomarker to assess myelin integrity affected by neurodegeneration and neuroinflamation processes as in Alzheimer's and Parkinson's disease or in multiple sclerosis \cite{faulkner2024harnessing}. Likewise, the intrinsic tissue susceptibility measured by QSM can probe the tissue iron content, which changes in several diseases such as Parkinson's, amyotrophic lateral sclerosis and multiple sclerosis \cite{eskreis2017clinical}. 

Hence, wu-MRSI enables multiparametric imaging of the brain by probing metabolites, tissue iron homeostasis and myelin integrity. wu-MRSI provides several potential advantages compared to sequential acquisitions: 1) time efficiency, by simultaneously imaging different biomarkers, 2) the multiparametric images are naturally coregistered, and 3) a simpler scan protocol. A limited number of studies\cite{peng2018simultaneous,guo2021simultaneous} have explored this possibility at a lower field (3T). Here, we demonstrate wu-MRSI at ultra-high field (7T) to take advantage of higher SNR, the larger spectral separation of metabolites peaks, the increased susceptibility anisotropy, and larger relaxation dispersion between free and bound water pools. To push the spatial resolution at 2 mm isotropic over the whole brain and accelerate the wu-MRSI we employed an ultra short-echo time with non-cartesian compressed-sense acquisition and a deep-learning based image reconstruction and spectral fitting. Compared to prior studies, our methodology achieves higher spatial resolution and larger brain coverage for metabolic imaging, while providing high enough resolution for QSM and MWF to probe tissue microstructure. We evaluated and compared our methodology to gold-standard methods in simulations, healthy controls and patients.

\section{Methods}\label{sec2}
\begin{figure*}[ht!] 
    \centering
    \includegraphics[width=1\textwidth]{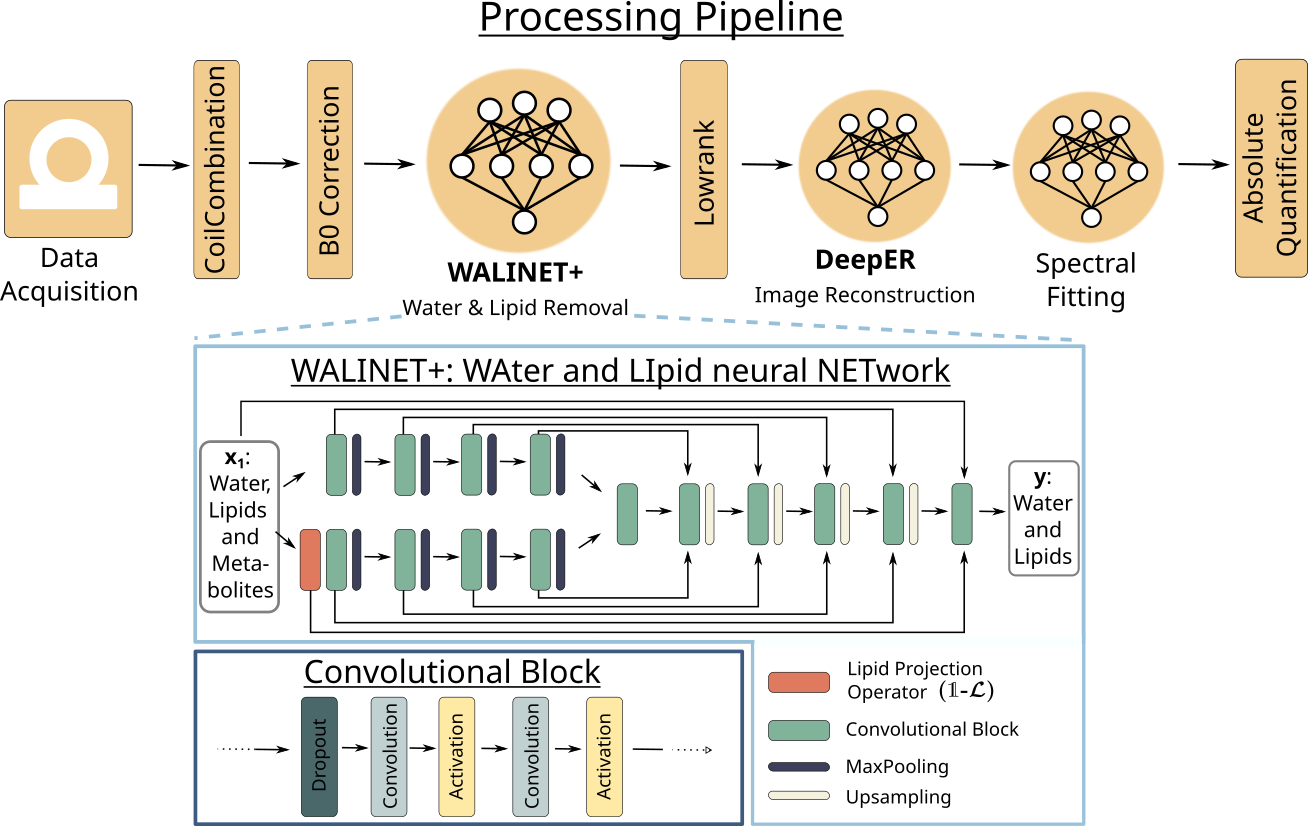} 
    \caption{End-to-end MRSI processing pipeline employing three deep neural networks: WALINET+ for nuisance signal removal, DeepER for (k,t) reconstruction, and spectral fitting  for metabolite quantification.}
    \label{fig:overview}
\end{figure*}

\textbf{Water-unsuppressed MRSI acquisition}
\newline
{In-vivo} MRSI data were acquired with 3D ${}^1$H-FID ECCENTRIC \cite{klauser2024eccentric} pulse sequences using a 7T MR scanner (MAGNETOM Terra.X, XA60 software, Siemens Healthineers, Erlangen, Germany) and a 1Tx/32Rx head coil (NovaMedical, Wilmington, MA, USA). The field of view (FOV) of $220 \times 220 \times 105$ mm$^3$, spectral bandwidth of 2280 Hz, and TE=0.9 ms were used for all protocols with a 1 ms SLR pulse of 6 kHz bandwidth and FA=27$^\circ$ for excitation. For the water-unsuppressed MRSI the acquisition used two water-unsuppressed protocols: a) TR=275 ms,  $64 \times 64 \times 31$ matrix, $3.4 \times 3.4 \times 3.4$ mm$^3$ isotropic voxel size, 451 time points, circle radius $0.25 k_{\max}$, compress-sense acceleration 2, with 9:21 minutes scan time, and b) TR=188 ms, $110 \times 110 \times 51$ matrix, $2 \times 2 \times 2$ mm$^3$ isotropic voxel size, 399 time points, circle radius $0.18 k_{\max}$, compress-sense acceleration 2, and 25:02 minutes scan time. No temporal interleaving was required for both protocols. The ultra high-resolution MRSI $2 \times 2 \times 2$ mm$^3$ data were retrospectively accelerated with compress-sense acceleration factor of 4 to 12:36 minutes scan time. For comparison, water-suppressed MRSI was acquired with the same protocol used for $3.4 \times 3.4 \times 3.4$ mm$^3$ wu-MRSI, and the water suppression was performed by including a 4-pulse WET block before the SLR pulse. The WET block adds 65 ms to the TR increasing the scan time by 33\%, so it was used only for acquiring the 3.4 mm ws-MRSI since it would have prolonged the acquisition of 2 mm ws-MRSI substantially. No fat suppression was applied in the sequence. B0 shimming was performed using the manufacturer's shimming tools that included four 3rd order shim coils which provided 30-40 Hz global linewidth of the water peak over the entire brain slab.   
The ECCENTRIC gradient waveforms with their smooth sinusoidal amplitude modulation reduce mechanical vibrations and water sidebands by two effects: 1) the bipolar oscillation of gradient amplitude creates opposite mechanical forces that partially compensate each other during ECCENTRIC readout, 2) the gradient amplitude and slew-rate required for ECCENTRIC trajectories are lower compared to other readout trajectories such as EPI for similar acquisition protocols, further reducing the mechanical forces and vibrations. Hence, the main source of vibrations for our ECCENTRIC pulse sequence are the spoilers at the end of the FID, which we optimized in the following way: 1) reduce the amplitude from 20 to 5 mT/m, 2) reduce the duration from 20 to 4ms, 3) reduce the slew rate by using a triangular shape instead of trapezoidal shape, and 4) alternate the polarity of the spoilers in consecutive TRs. These changes of the spoiler gradients can be performed safely considering that after 180 ms of the readout the signal decays by 2-3 orders of magnitude. We verified in phantoms that the new gradient spoilers destroy the signal at the end of the ECCENTRIC readout, which is sufficient for human scans considering that in-vivo signal decays faster than in phantoms. 

In addition to MRSI data, cartesian multi-echo GRE (ME-GRE) was acquired for ground-truth QSM and MWF imaging at the same spatial resolution following the consensus guidelines \cite{qsm2024recommended}. For QSM magnitude and phase images were acquired at 5 echo times: 4 ms, 10ms, 16 ms, 22 ms and 28 ms. For MWF magnitude images were acquired for 24 echoes times between 2 ms and 25 ms, with a 1 ms increment. Anatomical imaging was performed with MP2RAGE which was used to obtain quantitative T1 and PD maps for water, and in addition the FLAIR imaging was used for brain tumors.
\newline

\textbf{Water-unsuppressed MRSI processing.}
\newline
\newline
\textbf{\emph{WALINET+ strategy.}}
Similar to prior work \cite{weiser2025walinet}, a L2 lipid projection operator $(1-\mathcal{L})$, with $\mathcal{L}=(\mathbf{1}+\beta LL^H)^{-1}$ was derived for each subject from spectra located in the skull area. The operator is applied to the input spectrum $x_1$ containing water- $w$, sideband- $s$, lipid- $l$, and metabolite signal $m$.  
\begin{align}
    x_1 =& w+s+l+m \\
    x_2 =& (1-\mathcal{L}) x_1.
\end{align}
The WALINET+ neural network predicts the nuisance spectrum $y$ containing the nuisance signals $w+s+l$, which is subtracted from the input to derive the metabolites spectra.
\begin{align}
    m = x_1 - y
\end{align}

\textbf{\emph{Training data.}} 
\newline
Four types of data were used for training: 1) simulated metabolite spectra, 2) main water signal extracted by HLSVD from measured in-vivo wu-MRSI data, 3) water sidebands measured on water phantoms, 4) in-vivo fat signal extracted from wu-MRSI measured data. The simulated metabolites, the main water peak at 4.7 ppm and the lipid spectra were generated as in prior work \cite{weiser2025walinet}. Water signal extracted from water unsuppressed MRSI was used during the simulation of water suppressed data. During the simulation, the water signal was scaled with a random factor of up to $10^4$ times the metabolite signal. 

The water sidebands used for training were extracted from wu-MRSI data measured on water phantoms. The wu-MRSI data were measured with a variety of water phantoms, ECCENTRIC acquisition protocols including isocenter, off-isocenter, transverse and oblique slices. The water sidebands were obtained after removing the main water peak at 4.7 ppm using the HLSVD method. Randomized sideband signal augmentation included frequency shifts ($\pm60\text{Hz}$), mirroring along the center ($4.7\text{ppm}$), scaling of up to $1\%$ of the water amplitude and random phase shift.
Training data were generated from a cohort of 30 subjects (25 water-suppressed and 5 water-unsuppressed), including 5 patient cases. For each subject, 100,000 spectra were simulated, resulting in a total dataset of 3 million spectra. Data from two subjects were withheld for validation.
\newline

\textbf{\emph{Network architecture.}}
WALINET+ employs a Y-net architecture with 2 encoders and 1 decoder, depicted in Figure \ref{fig:overview}. Each encoder/decoder consists of 4 convolutional blocks followed by upsampling or maxpooling. Each block contains 2 convolutional layers and PReLU activation functions \cite{he2015delving}. Skip connections concatenate feature maps from each convolutional block with the decoder features on the same level. After the encoder, a convolutional block is employed as a connecting block. A kernel size of 3 is used for each convolutional layer, except for the final layer, which applies a kernel of size 1. The number of feature maps is doubled/halved after every block, except the first and last blocks mapping to 16 or 2 channels.
\newline

\textbf{\emph{Training details.}}
The simulated training metabolite spectra were augmented and normalized. The augmentation multiplies the input and ground truth spectra by the same random phase 
\begin{align}
    \phi = e^{i\omega}~ \omega \in [0,2\pi]
\end{align}
and scaling factor $s \in [0.5,1.5]$. 

Data normalization is performed by dividing the input spectra $x_1,x_2$ by the energy $E$.
\begin{align}
    E = \sqrt{|x_1-x_2|^T|x_1-x_2|}
\end{align}
The complex-valued spectra are separated into real and imaginary parts before being forwarded to the network.
Network training is performed by minimizing a mean-squared error loss between the ground-truth nuisance signal and the network estimation. We use the Adam optimizer \cite{kingma2014adam} with parameters $(\beta_1=0.9,\beta_2=0.999$ for first and second momentum updates. The network was trained for 4000 epochs on an NVIDIA Ampere A40 GPU with 48GB RAM.
\newline

\textbf{\emph{Alternative methods.}}
The WALINET+ was compared to several alternative methods: 1) WALINET, a deep learning approach for the removal of residual water and lipid signals from ws-MRSI, 2) a combination of HLSVD water suppression \cite{barkhuijsen1987improved} and L2-lipid regularization \cite{bilgic2014fast}, and 3) HLSVD+L2 and signal modulation for water sideband removal \cite{serrai2002localized, le2014fid}. 
For HLSVD water suppression, the 32 largest eigen-components of the spectra in the frequency range of $4.7 \pm 0.5\text{ppm}$ were used to identify the water signal, which was subtracted from the spectrum. L2 lipid regularization is performed by deriving a linear lipid-removal operator $\mathcal{L}$ from the spectra $L$ extracted from the skull mask.
\begin{align}
    \mathcal{L}=(\mathbf{1}+\beta LL^H)^{-1}
\end{align}
The regularization parameter $\beta$ is individually selected for each subject such that
\begin{align}
    \text{mean}(|\text{diag}(\mathcal{L})|) \sim 0.938.
\end{align}
This value was determined heuristically. The modulus method for water sideband removal \cite{serrai2002localized, le2014fid} was applied to the magnitude of the complex signal $s(t)$,
\begin{align}
    s(t)_{\textbf{mod}} = |s(t)|.
\end{align}
\newline

\textbf{\emph{End-to-end MRSI pipeline.}} 
In addition to WALINET+, the end-to-end deep learning MRSI processing pipeline included several other steps: 1) ESPIRIT coil combination \cite{uecker2014espirit}, 2) B0 correction, 3) low-rank model \cite{klauser2024eccentric}, 4) DeepER image reconstruction \cite{weiser2025deep}, 5) physics-based network for spectral fitting \cite{shamaei2023physics} and 6) absolute quantification \cite{baboli2024absolute}. Figure \ref{fig:overview} presents the pipeline diagram and the architecture of WALINET+.

Spectral fitting was performed with LCModel \cite{provencher1993estimation} for 3.4 mm MRSI data as a gold-standard for comparing wu-MRSI vs ws-MRSI. For the 2 mm MRSI data, a physics-based deep-learning model \cite{shamaei2023physics} was used for spectral fitting in order to speed-up metabolite fitting from 7 hours to few seconds. Both LCModel and the deep-learning model used a simulated basis set for brain metabolite spectra at TE=0.9 ms and 7T. Besides the normal brain metabolites, the 2-hydroxyglutarate (2HG) was included for isocitrate dehydrogenase (IDH) mutations in glioma tumors.

For absolute quantification, the metabolic maps were combined with the water maps obtained from wu-MRSI, and water T1 relaxation and proton density obtained from MP2RAGE imaging as described in \cite{baboli2024absolute}. For metabolite T1 relaxation, literature values at 7T were assumed \cite{xin2013proton}. Due to the very short echo time the T2 relaxation correction was not performed.   

The end-to-end deep-learning pipeline for MRSI processing is shown in Figure 1. The entire processing time for the 3.4 mm data is 7 min while for the 2 mm is 52 min on a PowerEdge R7525 server (Dell) with 64 CPU cores (AMD EPYC7542 2.90GHz, 128M Cache, DDR4-3200), 512 GB CPU RAM (RDIMM, 3200MT/s), 3 GPU NVIDIA Ampere A40 (PCIe, 48GB GPU RAM) running Rocky Linux release 8.8 (Green Obsidian).
\newline

\begin{figure*}[ht!] 
    \centering
    \includegraphics[width=\textwidth]{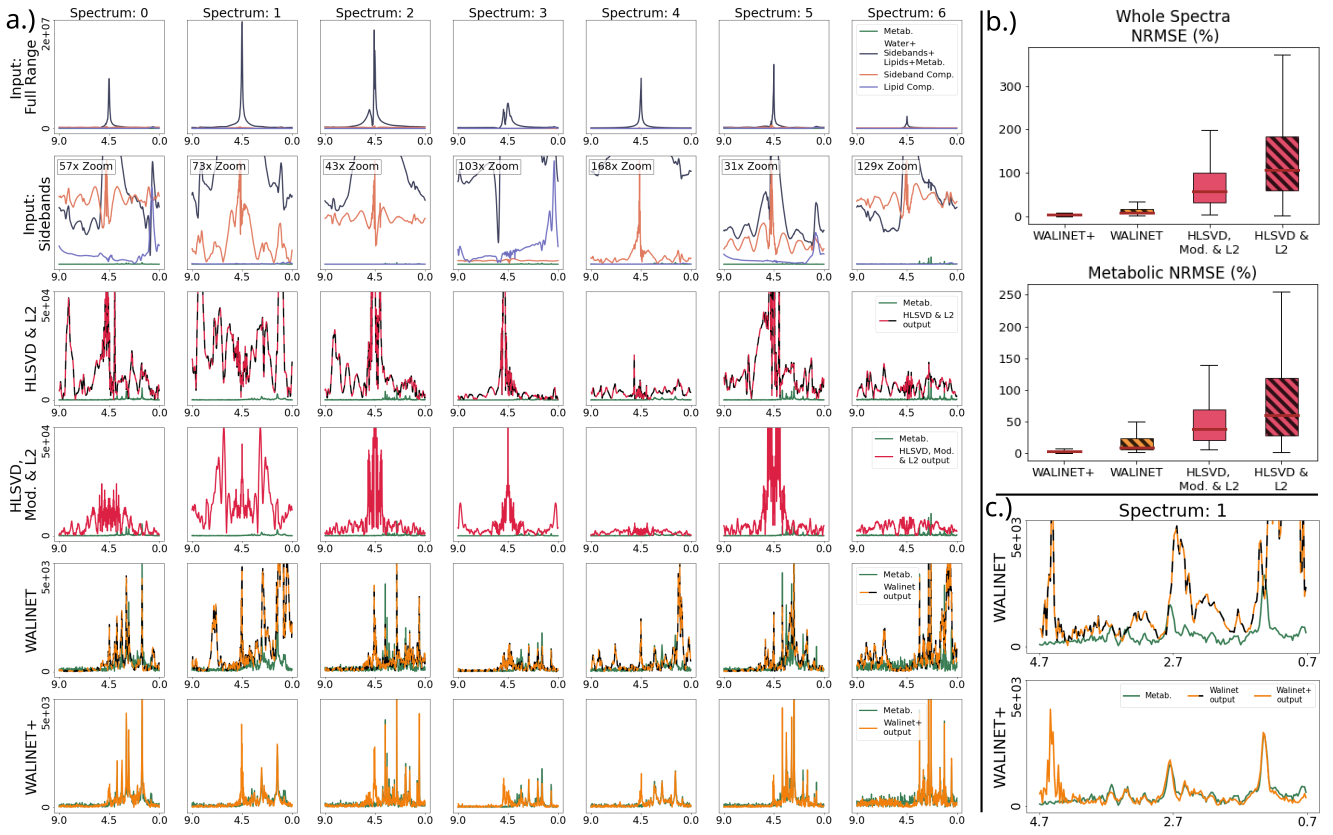} 
    \caption{WALINET+ results on simulated test data. We compare nuisance signal removal by WALINET+ with alternative methods such as WALINET, HLSVD+L2 and HSLVD modulus + L2. In a) examples of metabolite spectra contaminated with nuisance signals are shown at the top, and in the following rows after different nuisance removal methods. In b) we compare NRMSE between the ground truth metabolite signal and the recovered signal after nuisance removal. In c) zoomed spectra for WALINET and WALINET+.}
    \label{fig:simu}
\end{figure*}

\textbf{QSM and MWF imaging}
\newline
QSM reconstruction followed the processing pipeline as recommended by the consensus paper \cite{qsm2024recommended}. The first 56 echoes were extracted from the wu-MRSI dataset so that the last TE (25.06 ms) approximately matched the last TE of the standard multi-echo gradient-echo (GRE) data acquired for comparison (25 ms). A brain mask was generated using FSL BET \cite{smith2002fast}. Root sum of squares combination was computed across the echoes in the wu-MRSI data to create a magnitude volume, which was processed using the N4 bias correction algorithm to remove B1- bias field \cite{tustison2010n4itk}. Phase unwrapping was performed using ROMEO \cite{dymerska2021phase} and weighted echo combination was performed using the unwrapped echo phases assuming a constant tissue $T_2^*$  value of 25 ms. The phase quality map obtained from ROMEO was thresholded at 0.3 to refine the BET brain mask. The potential holes inside this refined mask were filled using the $imclose$ function. V-SHARP \cite{li2011quantitative} was used on the echo-combined, unwrapped phase to remove the background phase within the refined, hole-filled mask. V-SHARP yielded a yet smaller mask by eliminating regions at the mask boundary with unreliable convolution effects. For QSM, a final mask was created by re-introducing the holes coming from phase quality thresholding. Dipole inversion was performed using NDI with 500 iterations and L2 regularization weight of $1e-4$ \cite{polak2020nonlinear}. NDI used the N4 corrected magnitude and V-SHARP filtered tissue phase as input. Standard ME-GRE data was processed with the same pipeline to create susceptibility maps that were used for comparison.

Myelin water fraction (MWF) was estimated from the magnitude signal acquired with Cartesian multi-echo GRE and wu-MRSI. For GRE, a Tukey window (window size $= 0.4$) was applied to suppress ringing artifacts. For wu-MRSI, the echo train was cropped so that the last echo time matched that of GRE, ensuring consistent fitting conditions across sequences. The decay curve was modeled as the sum of fast-decaying (short $T_2^*$) and slow-decaying (long $T_2^*$) components, and MWF was defined as the fraction of the fast-decaying component. To enhance robustness to noise, a robust principal component analysis (rPCA)-based approach~\cite{song2020blind} $(\mu_1=1,\ \mu_2=1,\ \rho=0.5,\ \delta_1=0.01,\ \delta_2=0.01,\ \delta_3=0.01.\ \text{patch size}=10\times10\times10\ \text{voxels})$ was applied to both GRE and wu-MRSI for MWF mapping.
\newline

\textbf{Image metrics and statistical analysis}
\newline
The performance of methods for the removal of nuisance signal (main water peak, water sidebands, lipid signal)  was evaluated in terms of normalized root mean square error (NRMSE). To evaluate the quality of in-vivo MRSI data the signal-to-noise ratio (SNR), linewidth (FWHM), and Cramer-Rao lower bound (CRLB) were compared between wu-MRSI and ws-MRSI. Bland-Altman analyses were performed to compare metabolite levels obtained from ws-MRSI and wu-MRSI. In addition, Bland-Altman analyses were also performed to compare QSM and WMF obtained from ME-GRE and wu-MRSI acquisitions. 
\newline

\textbf{Human subjects}
A total of 30 subjects were scanned at the Athinoula A. Martinos Center for Biomedical Imaging with informed consent (Protocol 2013P001195), including 25
healthy volunteers (13M/12F, 21-49 years) and 5 patients with glioma tumors (3F/2M, 43-58 years).
\newline

\section{Results}\label{sec4}
WALINET+ performance was evaluated with the following strategy: 1) first, we investigated the performance of WALINET+ on simulated test data and compared it to alternative methods, 2) second, we compared the performance of WALINET+ on in-vivo wu-MRSI data vs the gold-standard ws-MRSI acquired with the high-resolution 3.4mm isotropic protocol, 3) third, we used WALINET+ on very high-resolution 2mm isotropic in-vivo wu-MRSI to simultaneously image metabolites. The ws-MRSI at 2mm was not performed because of significantly longer acquisition time associated with $33\%$ longer TR due to water suppression. QSM and MWF obtained from from wu-MRSI were compared to ground-truth ME-GRE.  

Figure \ref{fig:simu} shows the results obtained during the training and testing of WALINET+ on retrieving metabolite spectra from simulated data contaminated by the unsuppressed water peak, the water sidebands, and the lipid signals. Conventional methods such as HLSVD and the modulus HLSVD water removal combined with L2 lipid removal are able to clean the main water peak and most of lipids but large water sidebands are left which distort heavily the metabolite spectra. The WALINET network which was trained on water-suppressed data is able to clean better the spectra but water sidebands persist overlapping metabolite signals. The WALINET+ trained specifically for wu-MRSI is able to consistently remove all components of the nuissance signal and retrieve metabolites with high fidelity. Comparison of metabolite NRMSE indicate a mean of 91.8\% for HLSVD+L2, 54.8\% for HLSVD Modulus + L2, 17.9\% for WALINET and 3.4\% for WALINET+.

\begin{figure*}[ht!] 
    \centering
    \includegraphics[width=\textwidth]{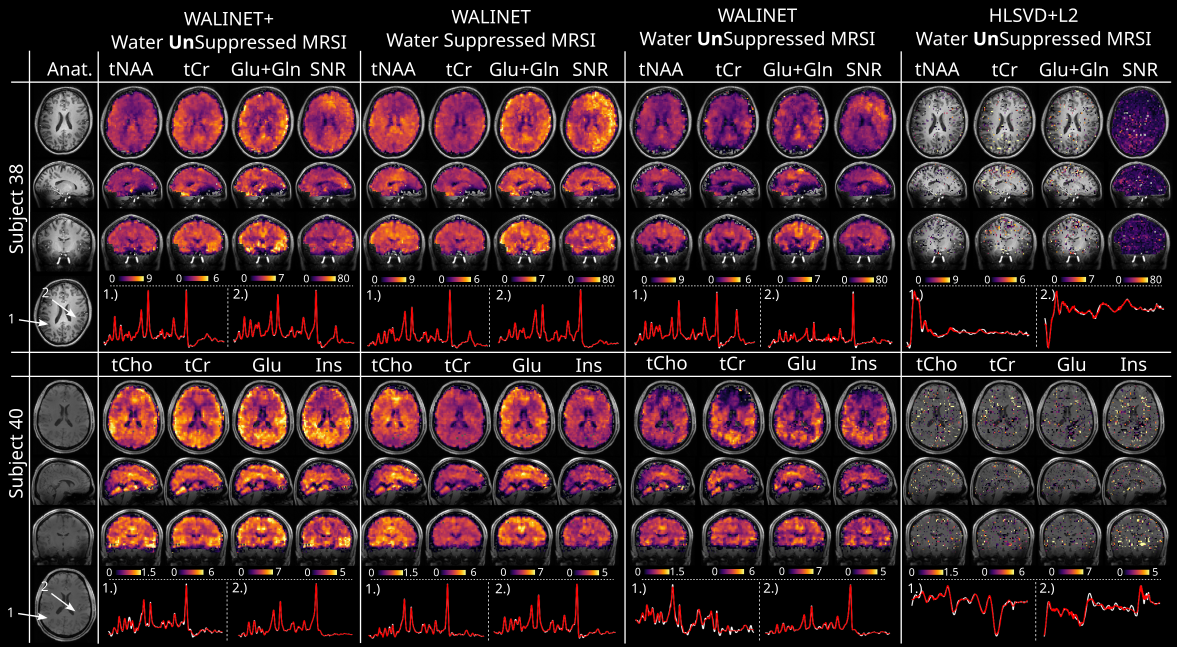} 
    \caption{In-vivo metabolic maps obtained by WALINET+ on wu-MRSI, compared to alternative methods for wu-MRSI and the ground-truth ws-MRSI. 7T MRSI data were acquired at 3.4 mm isotropic resolution in 9:20 minutes.}
    \label{fig:3dquali}
\end{figure*}

Figure \ref{fig:3dquali} shows in-vivo metabolic maps obtained in two healthy volunteers with the 3.4 mm protocol, comparing different methods for nuisance signal removal. The metabolic maps and spectra obtained by WALINET+ from wu-MRSI have similar spatial distribution as the ground-truth ws-MRSI. By contrast the use of WALINET and HLSVD+L2 on wu-MRSI results in significant differences in the maps and spectral patterns, with an almost complete failure of spectral fitting for HLSVD+L2 data. In particular, it can be seen that the creatine maps and the spectral peaks closer to the 4.7 ppm have higher signal in wu-MRSI than in the ws-MRSI, which is due to partial suppression of these signals in ws-MRSI by the WET water-suppression block.  

\begin{figure*}[ht!] 
    \centering
    \includegraphics[width=\textwidth]{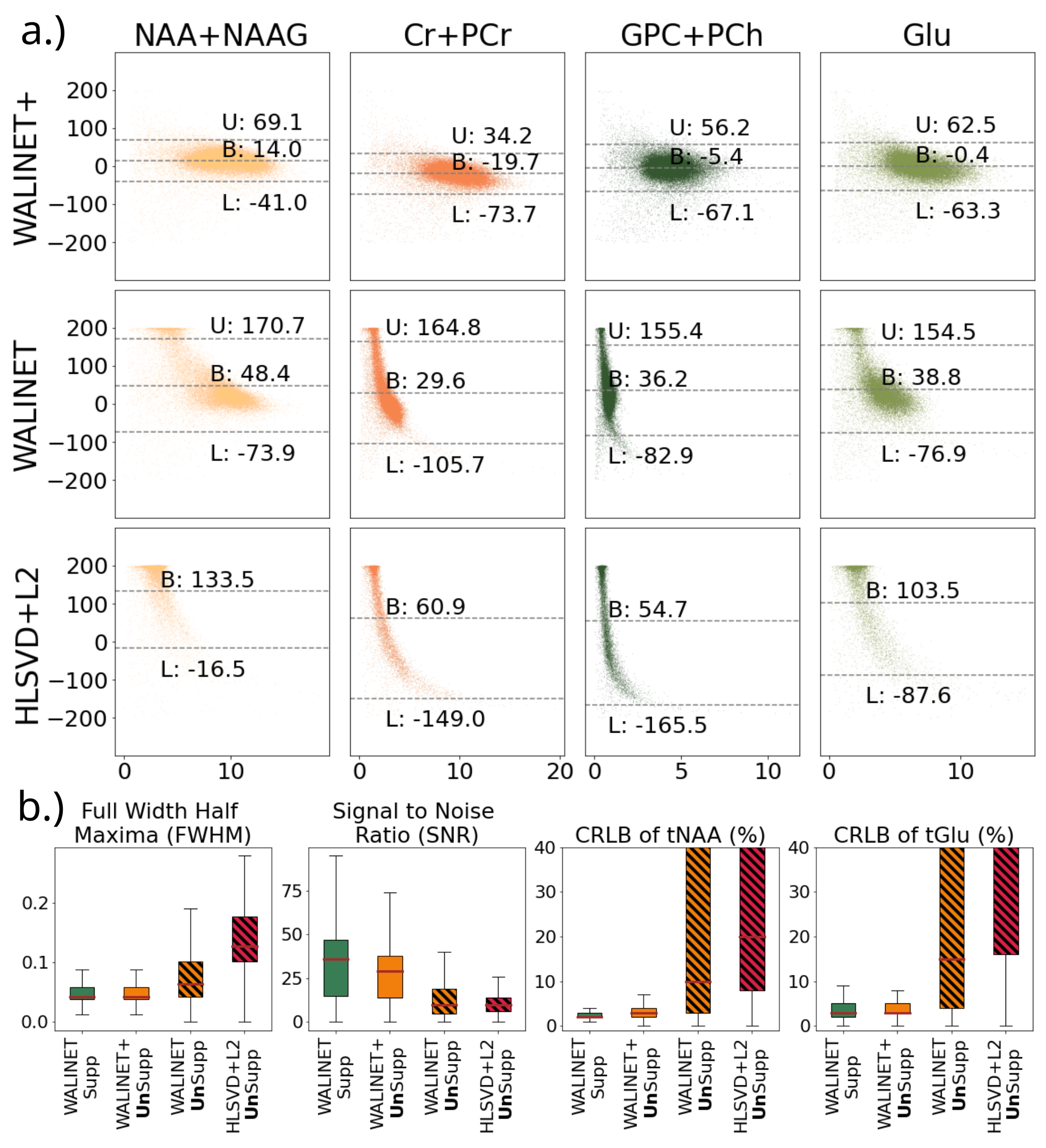} 
    \caption{Comparison of metabolite levels and spectral quality by in-vivo wu-MRSI and ws-MRSI. Bland-Altman plots are shown for metabolite levels in a), while boxplots for spectral quality are shown in b).}
    \label{fig:quant}
\end{figure*}

Figure \ref{fig:quant} provides a quantitative analysis of the different methods of nuisance signal removal on in-vivo MRSI data. The Bland-Altman plots of metabolite levels obtained by WALINET+ on wu-MRSI show the best agreement with ws-MRSI metabolite levels (bias $<20\%$, limits-of-agreement $\pm63\%$), while WALINET and HLSVD+L2 have larger bias and worse limits-of-agreement. Boxplots of spectral quality metrics show comparable quality between WALINET on ws-MRSI and WALINET+ on wu-MRSI with mean SNR of 34, mean FWHM of 0.041 ppm and relative CRLB of 3.3\%. By comparison much lower spectral quality is obtained by WALINET and HLSVD+L2 on wu-MRSI with mean SNR of 13.7, mean FWHM of 0.105 ppm and relative CRLB $>130\%$.

\begin{figure*}[ht!] 
    \centering
    \includegraphics[width=\textwidth]{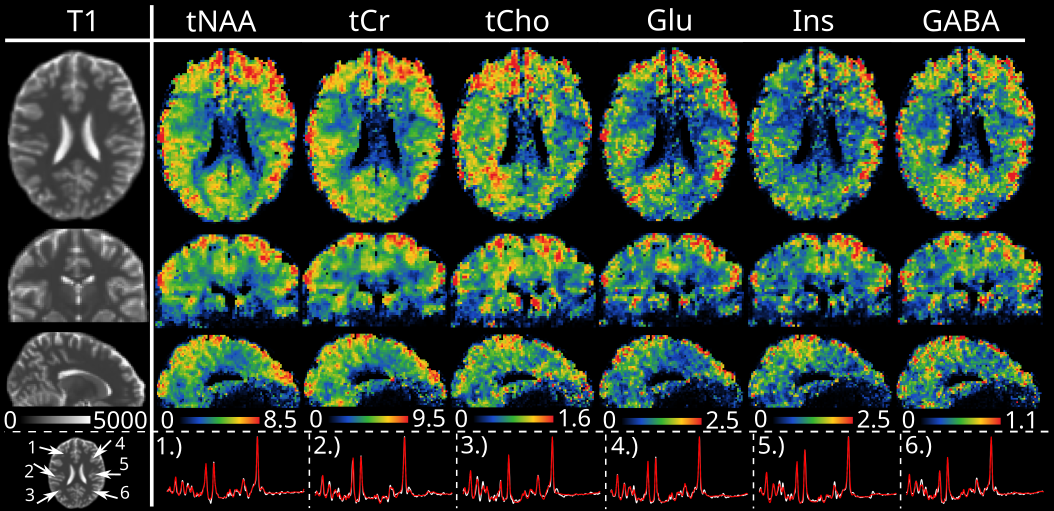} 
    \caption{Metabolic concentration [mM] maps obtained with wu-MRSI at 2 mm isotropic resolution from a healthy volunteer. ECCENTRIC acquisition was performed with a compressed-sense acceleration factor of 2, total acquisition time 25:02 minutes. The T1 map [ms] is shown on the left side. Arrows indicate the spectra location in the brain.}
    \label{fig:quali_vol46}
\end{figure*}

Figure \ref{fig:quali_vol46} shows 2 mm isotropic metabolic absolute concentration maps obtained by wu-MRSI in a healthy individual, using ECCENTRIC with a compress-sense acceleration factor of 2. Good contrast is observed between gray and white matter metabolic concentration that matches the structural features from the anatomical images. High spectral quality can be seen in the spectra throughout the brain. 

\begin{figure*}[ht!] 
    \centering
    \includegraphics[width=\textwidth]{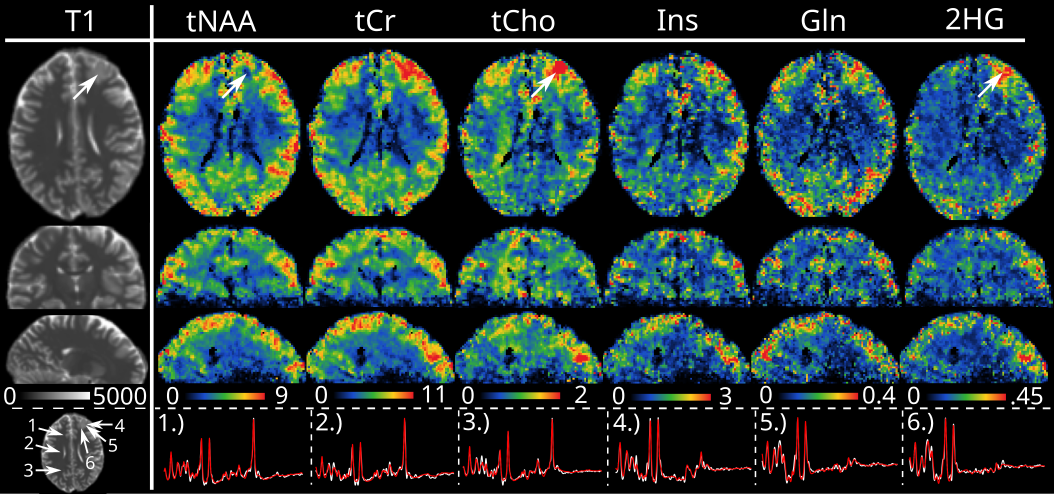} 
    \caption{Metabolic concentration [mM] maps obtained with wu-MRSI at 2 mm isotropic resolution from a mutant IDH glioma patient. ECCENTRIC acquisition was performed with a compressed-sense acceleration factor of 4, total acquisition time 12:36 minutes. Arrows in the upper row indicate the tumor location on the T1 map [ms] shown on the left side, and in the metabolic images. Arrows at the bottom show the spectra location in the healthy brain and tumor.}
    \label{fig:quali_pat13}
\end{figure*}

Figure \ref{fig:quali_pat13} shows 2 mm isotropic metabolic absolute concentration maps obtained with wu-MRSI in a mutant IDH glioma patient, using ECCENTRIC with a compress-sense acceleration factor of 4. The tumor is visible with increased concentration for total choline, total creatine, inositol, glutamine and 2-hydroxyglutarate, with a simultaneous reduction of total NAA. High quality spectra are obtained in the tumor as well the healthy brain.

\begin{figure*}[ht!] 
    \centering
    \includegraphics[width=\textwidth]{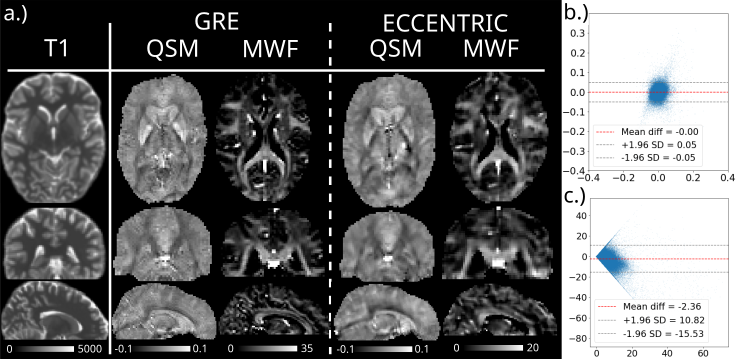} 
    \caption{QSM and MWF imaging in a healthy individual obtained from ws-MRSI acquisition with ECCENTRIC at 2 mm isotropic resolution compared to gold standard ME-GRE. On the right Bland-Altman plots compare the values of QSM (b) and MWF (c) methods. T1 map [ms] is shown on the left. }
    \label{fig:qsm_mwf_vol46}
\end{figure*}

\begin{figure*}[ht!] 
    \centering
    \includegraphics[width=\textwidth]{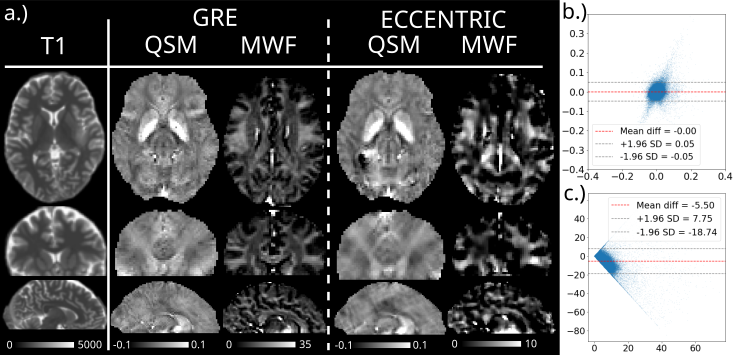} 
    \caption{QSM and MWF imaging in a mutant IDH glioma patient obtained from ws-MRSI acquisition with ECCENTRIC at 2 mm isotropic resolution compared to gold standard ME-GRE. On the right Bland-Altman plots compare the values of QSM (b) and MWF (c) methods. T1 map [ms] is shown on the left. }
    \label{fig:qsm_mwf_pat13}
\end{figure*}

Figures \ref{fig:qsm_mwf_vol46} and \ref{fig:qsm_mwf_pat13} show a comparison of QSM and MWF in one healthy volunteer (Fig 7) and one glioma patient (Fig 8) obtained with ME-GRE and wu-MRSI ECCENTRIC acquisition. Similar spatial patterns of tissue susceptibility anisotropy can be noticed in the QSM maps obtained by both methods. Likewise water myelin fraction shows a distribution that is consistent with the white matter spatial distribution in both methods. The Bland-Altman comparison between the two methods for the quantitative values of tissue susceptibility and water myelin fraction reveals a very small bias and narrow limits of agreement.

\section{Discussion}\label{sec3}
Metabolic MRSI, QSM and MWF benefit from ultra-high field due to increased spectral separation, larger susceptibility anisotropy, and wider relaxation dispersion, respectively, as well as higher SNR. In addition, the multi-parametric imaging benefits from simultaneous data acquisition by reducing the total scan time, minimizing protocol complexity and providing images that are naturally coregistered. 

In this work we harnessed the opportunities of ultra-high field by developing water-unsuppressed MRSI to simultaneously image metabolite and water signals. The overwhelming unsuppressed water signal poses major technical challenges for metabolite quantification, especially through the gradient induced water sidebands that overlap metabolite peaks. For this we developed a new processing method WALINET+ for water sideband removal based on a deep-learning approach. To the best of our knowledge, WALINET+ is the first demonstration of deep-learning water sideband removal from wu-MRSI. WALINET+ provides several advantages: it 1) eliminates the need of complex physical modeling that requires intimate characterization of the scanner hardware performance, 2) allows the use of all the time points in ultra-short echo time FID acquisitions, 3) eliminates the need of water based suppression techniques such as WET that prolong TR, increase SAR and are susceptible to B0 and B1 inhomogeneity at ultra-high field, 4) eliminates the need of metabolite cycling methods that double the scan time and increase SAR. WALINET+ learns the water sidebands directly from the wu-MRSI data and can be trained with a large amount of spectra from high-resolution wu-MRSI acquisitions under various conditions and protocols. The large amount of training data spans a comprehensive diverse distribution of sidebands acquired with different gradient trajectories and off-isocenter positions, leading to robust sideband removal. Furthermore, we combined the removal of three types of nuisance signals (main water peak, water sidebands and lipid signals) into a single step, which are conventionally handled in separate processing steps. Our approach is the first method that unifies these tasks.     

Compared to prior wu-MRSI shown at 3T\cite{peng2018simultaneous,guo2021simultaneous} for simultaneous metabolic, QSM and MWF imaging our study has several distinctive features. In particular, related to metabolic imaging: 1) we extend the processing pipeline to obtain absolute concentration maps for metabolites, 2) we obtain higher resolution metabolic images, and 3) we obtain larger brain slab coverage. On the other hand, our current implementation has several limitations by providing lower resolution for WMF and QSM compared to prior simultaneous MRSI-QSM-MWF at 3T and longer scan times. Our fastest acquisition time of 12 min for 2 mm ECCENTRIC wu-MRSI is double compared to the time required by SPICE\cite{peng2018simultaneous,guo2021simultaneous} at 3T that had lower resolution  (3 mm isotropic) for metabolic imaging but higher in-plane resolution (1x1x1.9 $mm^3$) for QSM and MWF. 

Although we present some of the highest spatial resolution metabolic imaging obtained at 7T for near whole-brain coverage, the limitation of our methodology is represented by the lower spatial resolution of QSM and MWF compared to conventional ME-GRE. However, further acceleration is possible with ECCENTRIC and we plan to investigate this in the future to speed-up the acquisition and obtain higher spatial resolution imaging.

Our end-to-end processing pipeline of wu-MRSI data for metabolic imaging integrates three neural networks (WALINET+, DeepER, and spectral fitting) which allows significant speed-up of the reconstruction times to 7 min for the 3.4 mm protocol and 52 min for the 2 mm protocol compared to 10 hours and 2 days, respectively, by conventional methods.   

The results on healthy volunteers and glioma patients show good agreement between wu-MRSI and gold standard ws-MRSI or ME-GRE for metabolic imaging or QSM and MWF, respectively. In particular, we noticed that the spectral linewidth improves significantly at 2 mm ultra-high resolution, which reduces spectral overlap and helps with fitting important metabolites such as 2-hydroxyglurate (2HG) in mutant IDH glioma patients \cite{esmaeili2020integrated}. 

Few differences can be noticed between wu-MRSI and the gold standard methods: 1) several metabolites such as creatine and glutamate have higher SNR in wu-MRSI compared to ws-MRSI due to the fact that water suppression reduces the signal of these metabolites either by direct effect of the water-suppression RF pulses or by the chemical-exchange saturation effect, 2) the QSM and MWF derived from wu-MRSI have slightly less sharp edges around anatomical landmarks compared to the ones derived from the ME-GRE. The latter may be due to blurring caused by head motion during the longer acquisition time of the wu-MRSI compared the ME-GRE. Incorporation of motion-correction strategies \cite{bogner20143d}, shortening the acquisition by more acceleration \cite{guo2021simultaneous}  and super-resolution reconstruction \cite{li2020super} could help improve the structural features of QSM and MWF derived from wu-MRSI.    

In conclusion, our work demonstrates the feasibility of simultaneous quantitative metabolic, susceptibility and myelin water imaging at ultra-high field. This has been achieved by combining a fast high-resolution wu-MRSI acquisition by non-cartesian compressed-sense ECCENTRIC encoding and an end-to-end deep learning pipeline that combines robust water sideband removal, rapid (k,t) reconstruction and fast spectral fitting. We anticipate that our methodology will help advance research into human brain function and structure in health and disease.      

\section*{Funding}
This work was supported by the National Institutes of Health
(2R01CA211080-06A1, R01CA255479, P50CA165962, P41EB015896, R00HD101553),
the Harvard Brain Initiative Bipolar Disorder Seed Grant,
and the Athinoula A. Martinos Center for Biomedical Imaging.

\printbibliography

@article{song2020blind,
  title={Blind source separation for myelin water fraction mapping using multi-echo gradient echo imaging},
  author={Song, Jae Eun and Shin, Jaewook and Lee, Hongpyo and Lee, Ho Joon and Moon, Won-Jin and Kim, Dong-Hyun},
  journal={IEEE Transactions on Medical Imaging},
  volume={39},
  number={6},
  pages={2235--2245},
  year={2020},
  publisher={IEEE}
}

@article{tkac2005methodology,
  title={Methodology of1H NMR spectroscopy of the human brain at very high magnetic fields},
  author={Tk{\'a}{\'c}, I and Gruetter, R},
  journal={Applied magnetic resonance},
  volume={29},
  number={1},
  pages={139--157},
  year={2005},
  publisher={Springer}
}

@article{tkavc1999vivo,
  title={In vivo 1H NMR spectroscopy of rat brain at 1 ms echo time},
  author={Tk{\'a}{\v{c}}, Ivan and Star{\v{c}}uk, Z and Choi, I-Y and Gruetter, R},
  journal={Magnetic Resonance in Medicine: An Official Journal of the International Society for Magnetic Resonance in Medicine},
  volume={41},
  number={4},
  pages={649--656},
  year={1999},
  publisher={Wiley Online Library}
}

@article{tkavc2001vivo,
  title={In vivo 1H NMR spectroscopy of the human brain at 7 T},
  author={Tk{\'a}{\v{c}}, Ivan and Andersen, Peter and Adriany, Gregor and Merkle, Hellmut and Uǧurbil, K{\^a}mil and Gruetter, Rolf},
  journal={Magnetic Resonance in Medicine: An Official Journal of the International Society for Magnetic Resonance in Medicine},
  volume={46},
  number={3},
  pages={451--456},
  year={2001},
  publisher={Wiley Online Library}
}

@article{ogg1994wet,
  title={WET, a T1-and B1-insensitive water-suppression method for in vivo localized 1H NMR spectroscopy},
  author={Ogg, Robert J and Kingsley, RB and Taylor, June S},
  journal={Journal of Magnetic Resonance, Series B},
  volume={104},
  number={1},
  pages={1--10},
  year={1994},
  publisher={Elsevier}
}

@article{barkhuijsen1987improved,
  title={Improved algorithm for noniterative time-domain model fitting to exponentially damped magnetic resonance signals},
  author={Barkhuijsen, H and De Beer, R and Van Ormondt, D},
  journal={Journal of magnetic resonance (1969)},
  volume={73},
  number={3},
  pages={553--557},
  year={1987},
  publisher={Elsevier}
}

@article{weiser2025walinet,
  title={WALINET: A water and lipid identification convolutional neural network for nuisance signal removal in 1 H\^{} 1 H MR spectroscopic imaging},
  author={Weiser, Paul J and Langs, Georg and Motyka, Stanislav and Bogner, Wolfgang and Courvoisier, S{\'e}bastien and Hoffmann, Malte and Klauser, Antoine and Andronesi, Ovidiu C},
  journal={Magnetic Resonance in Medicine},
  volume={93},
  number={4},
  pages={1430--1442},
  year={2025},
  publisher={Wiley Online Library}
}

@article{le2014fid,
  title={FID modulus: a simple and efficient technique to phase and align MR spectra},
  author={Le Fur, Yann and Cozzone, Patrick J},
  journal={Magnetic Resonance Materials in Physics, Biology and Medicine},
  volume={27},
  number={2},
  pages={131--148},
  year={2014},
  publisher={Springer}
}

@article{serrai2002localized,
  title={Localized proton spectroscopy without water suppression: removal of gradient induced frequency modulations by modulus signal selection},
  author={Serrai, Ha{\c{c}}{\`e}ne and Clayton, David B and Senhadji, Lotfi and Zuo, Chun and Lenkinski, Robert E},
  journal={Journal of magnetic resonance},
  volume={154},
  number={1},
  pages={53--59},
  year={2002},
  publisher={Elsevier}
}

@article{chang2018non,
  title={Non-water-suppressed 1H FID-MRSI at 3T and 9.4 T},
  author={Chang, Paul and Nassirpour, Sahar and Avdievitch, Nikolai and Henning, Anke},
  journal={Magnetic resonance in medicine},
  volume={80},
  number={2},
  pages={442--451},
  year={2018},
  publisher={Wiley Online Library}
}

@article{tkavc2021water,
  title={Water and lipid suppression techniques for advanced 1H MRS and MRSI of the human brain: experts' consensus recommendations},
  author={Tk{\'a}{\v{c}}, Ivan and Deelchand, Dinesh and Dreher, Wolfgang and Hetherington, Hoby and Kreis, Roland and Kumaragamage, Chathura and Pova{\v{z}}an, Michal and Spielman, Daniel M and Strasser, Bernhard and de Graaf, Robin A},
  journal={NMR in Biomedicine},
  volume={34},
  number={5},
  pages={e4459},
  year={2021},
  publisher={Wiley Online Library}
}

@article{kingma2014adam,
  title={Adam: A method for stochastic optimization},
  author={Kingma, Diederik P and Ba, Jimmy},
  journal={arXiv preprint arXiv:1412.6980},
  year={2014}
}

@article{bilgic2014fast,
  title={Fast image reconstruction with L2-regularization},
  author={Bilgic, Berkin and Chatnuntawech, Itthi and Fan, Audrey P and Setsompop, Kawin and Cauley, Stephen F and Wald, Lawrence L and Adalsteinsson, Elfar},
  journal={Journal of magnetic resonance imaging},
  volume={40},
  number={1},
  pages={181--191},
  year={2014},
  publisher={Wiley Online Library}
}

@inproceedings{he2015delving,
  title={Delving deep into rectifiers: Surpassing human-level performance on imagenet classification},
  author={He, Kaiming and Zhang, Xiangyu and Ren, Shaoqing and Sun, Jian},
  booktitle={Proceedings of the IEEE international conference on computer vision},
  pages={1026--1034},
  year={2015}
}

@article{qsm2024recommended,
  title={Recommended implementation of quantitative susceptibility mapping for clinical research in the brain: a consensus of the ISMRM electro-magnetic tissue properties study group},
  author={QSM Consensus Organization Committee and Bilgic, Berkin and Costagli, Mauro and Chan, Kwok-Shing and Duyn, Jeff and Langkammer, Christian and Lee, Jongho and Li, Xu and Liu, Chunlei and Marques, Jos{\'e} P and others},
  journal={Magnetic resonance in medicine},
  volume={91},
  number={5},
  pages={1834--1862},
  year={2024},
  publisher={Wiley Online Library}
}

@article{smith2002fast,
  title={Fast robust automated brain extraction},
  author={Smith, Stephen M},
  journal={Human brain mapping},
  volume={17},
  number={3},
  pages={143--155},
  year={2002},
  publisher={Wiley Online Library}
}

@article{tustison2010n4itk,
  title={N4ITK: improved N3 bias correction},
  author={Tustison, Nicholas J and Avants, Brian B and Cook, Philip A and Zheng, Yuanjie and Egan, Alexander and Yushkevich, Paul A and Gee, James C},
  journal={IEEE transactions on medical imaging},
  volume={29},
  number={6},
  pages={1310--1320},
  year={2010},
  publisher={IEEE}
}

@article{dymerska2021phase,
  title={Phase unwrapping with a rapid opensource minimum spanning tree algorithm (ROMEO)},
  author={Dymerska, Barbara and Eckstein, Korbinian and Bachrata, Beata and Siow, Bernard and Trattnig, Siegfried and Shmueli, Karin and Robinson, Simon Daniel},
  journal={Magnetic resonance in medicine},
  volume={85},
  number={4},
  pages={2294--2308},
  year={2021},
  publisher={Wiley Online Library}
}

@article{li2011quantitative,
  title={Quantitative susceptibility mapping of human brain reflects spatial variation in tissue composition},
  author={Li, Wei and Wu, Bing and Liu, Chunlei},
  journal={Neuroimage},
  volume={55},
  number={4},
  pages={1645--1656},
  year={2011},
  publisher={Elsevier}
}

@article{polak2020nonlinear,
  title={Nonlinear dipole inversion (NDI) enables robust quantitative susceptibility mapping (QSM)},
  author={Polak, Daniel and Chatnuntawech, Itthi and Yoon, Jaeyeon and Iyer, Siddharth Srinivasan and Milovic, Carlos and Lee, Jongho and Bachert, Peter and Adalsteinsson, Elfar and Setsompop, Kawin and Bilgic, Berkin},
  journal={NMR in Biomedicine},
  volume={33},
  number={12},
  pages={e4271},
  year={2020},
  publisher={Wiley Online Library}
}

@article{peng2018simultaneous,
  title={Simultaneous QSM and metabolic imaging of the brain using SPICE},
  author={Peng, Xi and Lam, Fan and Li, Yudu and Clifford, Bryan and Liang, Zhi-Pei},
  journal={Magnetic resonance in medicine},
  volume={79},
  number={1},
  pages={13--21},
  year={2018},
  publisher={Wiley Online Library}
}

@article{baboli2024absolute,
  title={Absolute metabolite quantification in individuals with glioma and healthy individuals using whole-brain three-dimensional MR spectroscopic and echo-planar time-resolved imaging},
  author={Baboli, Mehran and Wang, Fuyixue and Dong, Zijing and Dietrich, Jorg and Uhlmann, Erik J and Batchelor, Tracy T and Cahill, Daniel P and Andronesi, Ovidiu C},
  journal={Radiology},
  volume={312},
  number={3},
  pages={e232401},
  year={2024},
  publisher={Radiological Society of North America}
}

@article{xin2013proton,
  title={Proton T1 relaxation times of metabolites in human occipital white and gray matter at 7 T},
  author={Xin, Lijing and Schaller, Beno{\^\i}t and Mlynarik, Vladimir and Lu, Huanxiang and Gruetter, Rolf},
  journal={Magnetic resonance in medicine},
  volume={69},
  number={4},
  pages={931--936},
  year={2013},
  publisher={Wiley Online Library}
}

@article{chadzynski2010chemical,
  title={Chemical shift imaging without water suppression at 3 T},
  author={Chadzynski, Grzegorz L and Klose, Uwe},
  journal={Magnetic Resonance Imaging},
  volume={28},
  number={5},
  pages={669--675},
  year={2010},
  publisher={Elsevier}
}

@article{nixon2008compensation,
  title={Compensation of gradient-induced magnetic field perturbations},
  author={Nixon, Terence W and McIntyre, Scott and Rothman, Douglas L and de Graaf, Robin A},
  journal={Journal of Magnetic Resonance},
  volume={192},
  number={2},
  pages={209--217},
  year={2008},
  publisher={Elsevier}
}

@article{guo2021simultaneous,
  title={Simultaneous QSM and metabolic imaging of the brain using SPICE: Further improvements in data acquisition and processing},
  author={Guo, Rong and Zhao, Yibo and Li, Yudu and Wang, Tianyao and Li, Yao and Sutton, Brad and Liang, Zhi-Pei},
  journal={Magnetic resonance in medicine},
  volume={85},
  number={2},
  pages={970--977},
  year={2021},
  publisher={Wiley Online Library}
}

@article{andronesi2018pharmacodynamics,
  title={Pharmacodynamics of mutant-IDH1 inhibitors in glioma patients probed by in vivo 3D MRS imaging of 2-hydroxyglutarate},
  author={Andronesi, Ovidiu C and Arrillaga-Romany, Isabel C and Ly, K Ina and Bogner, Wolfgang and Ratai, Eva M and Reitz, Kara and Iafrate, A John and Dietrich, Jorg and Gerstner, Elizabeth R and Chi, Andrew S and others},
  journal={Nature communications},
  volume={9},
  number={1},
  pages={1474},
  year={2018},
  publisher={Nature Publishing Group UK London}
}

@article{andronesi2020imaging,
  title={Imaging neurochemistry and brain structure tracks clinical decline and mechanisms of ALS in patients},
  author={Andronesi, Ovidiu C and Nicholson, Katharine and Jafari-Khouzani, Kourosh and Bogner, Wolfgang and Wang, Jing and Chan, James and Macklin, Eric A and Levine-Weinberg, Mark and Breen, Christopher and Schwarzschild, Michael A and others},
  journal={Frontiers in neurology},
  volume={11},
  pages={590573},
  year={2020},
  publisher={Frontiers Media SA}
}

@article{oz2014clinical,
  title={Clinical proton MR spectroscopy in central nervous system disorders},
  author={{\"O}z, G{\"u}lin and Alger, Jeffry R and Barker, Peter B and Bartha, Robert and Bizzi, Alberto and Boesch, Chris and Bolan, Patrick J and Brindle, Kevin M and Cudalbu, Cristina and Din{\c{c}}er, Alp and others},
  journal={Radiology},
  volume={270},
  number={3},
  pages={658--679},
  year={2014},
  publisher={Radiological Society of North America}
}

@article{dong2015proton,
  title={Proton MRS and MRSI of the brain without water suppression},
  author={Dong, Zhengchao},
  journal={Progress in nuclear magnetic resonance spectroscopy},
  volume={86},
  pages={65--79},
  year={2015},
  publisher={Elsevier}
}

@article{dreher2005new,
  title={New method for the simultaneous detection of metabolites and water in localized in vivo 1H nuclear magnetic resonance spectroscopy},
  author={Dreher, Wolfgang and Leibfritz, Dieter},
  journal={Magnetic Resonance in Medicine: An Official Journal of the International Society for Magnetic Resonance in Medicine},
  volume={54},
  number={1},
  pages={190--195},
  year={2005},
  publisher={Wiley Online Library}
}

@article{koehl1999linear,
  title={Linear prediction spectral analysis of NMR data},
  author={Koehl, P},
  journal={Progress in Nuclear Magnetic Resonance Spectroscopy},
  volume={34},
  number={3-4},
  pages={257--299},
  year={1999},
  publisher={Elsevier}
}

@article{faulkner2024harnessing,
  title={Harnessing myelin water fraction as an imaging biomarker of human cerebral aging, neurodegenerative diseases, and risk factors influencing myelination: A review},
  author={Faulkner, Mary E and Gong, Zhaoyuan and Guo, Alex and Laporte, John P and Bae, Jonghyun and Bouhrara, Mustapha},
  journal={Journal of neurochemistry},
  volume={168},
  number={9},
  pages={2243--2263},
  year={2024},
  publisher={Wiley Online Library}
}

@article{eskreis2017clinical,
  title={The clinical utility of QSM: disease diagnosis, medical management, and surgical planning},
  author={Eskreis-Winkler, Sarah and Zhang, Yan and Zhang, Jingwei and Liu, Zhe and Dimov, Alexey and Gupta, Ajay and Wang, Yi},
  journal={NMR in Biomedicine},
  volume={30},
  number={4},
  pages={e3668},
  year={2017},
  publisher={Wiley Online Library}
}

@article{uecker2014espirit,
  title={ESPIRiT—an eigenvalue approach to autocalibrating parallel MRI: where SENSE meets GRAPPA},
  author={Uecker, Martin and Lai, Peng and Murphy, Mark J and Virtue, Patrick and Elad, Michael and Pauly, John M and Vasanawala, Shreyas S and Lustig, Michael},
  journal={Magnetic resonance in medicine},
  volume={71},
  number={3},
  pages={990--1001},
  year={2014},
  publisher={Wiley Online Library}
}

@article{weiser2025deep,
  title={Deep-ER: Deep Learning ECCENTRIC Reconstruction for fast high-resolution neurometabolic imaging},
  author={Weiser, Paul J and Langs, Georg and Bogner, Wolfgang and Motyka, Stanislav and Strasser, Bernhard and Golland, Polina and Singh, Nalini and Dietrich, Jorg and Uhlmann, Erik and Batchelor, Tracy and others},
  journal={NeuroImage},
  volume={309},
  pages={121045},
  year={2025},
  publisher={Elsevier}
}

@article{shamaei2023physics,
  title={Physics-informed deep learning approach to quantification of human brain metabolites from magnetic resonance spectroscopy data},
  author={Shamaei, Amirmohammad and Starcukova, Jana and Starcuk Jr, Zenon},
  journal={Computers in Biology and Medicine},
  volume={158},
  pages={106837},
  year={2023},
  publisher={Elsevier}
}

@article{provencher1993estimation,
  title={Estimation of metabolite concentrations from localized in vivo proton NMR spectra},
  author={Provencher, Stephen W},
  journal={Magnetic resonance in medicine},
  volume={30},
  number={6},
  pages={672--679},
  year={1993},
  publisher={Wiley Online Library}
}

@article{esmaeili2020integrated,
  title={An integrated RF-receive/B0-shim array coil boosts performance of whole-brain MR spectroscopic imaging at 7 T},
  author={Esmaeili, Morteza and Stockmann, Jason and Strasser, Bernhard and Arango, Nicolas and Thapa, Bijaya and Wang, Zhe and van der Kouwe, Andre and Dietrich, Jorg and Cahill, Daniel P and Batchelor, Tracy T and others},
  journal={Scientific Reports},
  volume={10},
  number={1},
  pages={15029},
  year={2020},
  publisher={Nature Publishing Group UK London}
}

@article{li2020super,
  title={Super-resolution whole-brain 3D MR spectroscopic imaging for mapping d-2-hydroxyglutarate and tumor metabolism in isocitrate dehydrogenase 1--mutated human gliomas},
  author={Li, Xianqi and Strasser, Bernhard and Jafari-Khouzani, Kourosh and Thapa, Bijaya and Small, Julia and Cahill, Daniel P and Dietrich, Jorg and Batchelor, Tracy T and Andronesi, Ovidiu C},
  journal={Radiology},
  volume={294},
  number={3},
  pages={589--597},
  year={2020},
  publisher={Radiological Society of North America}
}

@article{bogner20143d,
  title={3D GABA imaging with real-time motion correction, shim update and reacquisition of adiabatic spiral MRSI},
  author={Bogner, Wolfgang and Gagoski, Borjan and Hess, Aaron T and Bhat, Himanshu and Tisdall, M Dylan and van der Kouwe, Andre JW and Strasser, Bernhard and Marja{\'n}ska, Ma{\l}gorzata and Trattnig, Siegfried and Grant, Ellen and others},
  journal={Neuroimage},
  volume={103},
  pages={290--302},
  year={2014},
  publisher={Elsevier}
}

@article{klauser2024eccentric,
  title={ECCENTRIC: a fast and unrestrained approach for high-resolution in vivo metabolic imaging at ultra-high field MR},
  author={Klauser, Antoine and Strasser, Bernhard and Bogner, Wolfgang and Hingerl, Lukas and Courvoisier, Sebastien and Schirda, Claudiu and Rosen, Bruce R and Lazeyras, Francois and Andronesi, Ovidiu C},
  journal={Imaging Neuroscience},
  volume={2},
  pages={1--20},
  year={2024},
  publisher={MIT Press 255 Main Street, 9th Floor, Cambridge, Massachusetts 02142, USA~…}
}

@article{bilgic2013lipid,
  title={Lipid suppression in CSI with spatial priors and highly undersampled peripheral k-space},
  author={Bilgic, Berkin and Gagoski, Borjan and Kok, Trina and Adalsteinsson, Elfar},
  journal={Magnetic resonance in medicine},
  volume={69},
  number={6},
  pages={1501--1511},
  year={2013},
  publisher={Wiley Online Library}
}

@article{heckova2022ms,
  title={Extensive Brain Pathologic Alterations Detected with 7.0-T MR Spectroscopic Imaging Associated with Disability in Multiple Sclerosis},
  author={Heckova E and Dal-Bianco A and Strasser B and Hangel GJ and Lipka A and Motyka S and Hingerl L and Rommer PS and Berger T and Hnilicová P and Kantorová E and Leutmezer F and Kurča E and Gruber S and Trattnig S and Bogner W},
  journal={Radiology},
  volume={303},
  number={1},
  pages={141--150},
  year={2022}
}

@article{hu2024ad,
  title={Neurometabolic topography and associations with cognition in Alzheimer's disease: A whole-brain high-resolution 3D MRSI study },
author={Hu J and Zhang M and Zhang Y and Zhuang H and Zhao Y and Li Y and Jin W and Qian XH and Wang L and Ye G and Tang H and Liu J and Li B and Nachev P and Liang ZP and Li Y.},
  journal={Alzheimers Dementia},
  volume={20},
  number={9},
  pages={6407--6422},
  year={2024}
}

@article{ma2016removal,
  title={Removal of nuisance signals from limited and sparse 1H MRSI data using a union-of-subspaces model},
  author={Ma, Chao and Lam, Fan and Johnson, Curtis L and Liang, Zhi-Pei},
  journal={Magnetic resonance in medicine},
  volume={75},
  number={2},
  pages={488--497},
  year={2016},
  publisher={Wiley Online Library}
}

@article{lin2019water,
  title={Water removal in MR spectroscopic imaging with L2 regularization},
  author={Lin, Liangjie and Pova{\v{z}}an, Michal and Berrington, Adam and Chen, Zhong and Barker, Peter B},
  journal={Magnetic resonance in medicine},
  volume={82},
  number={4},
  pages={1278--1287},
  year={2019},
  publisher={Wiley Online Library}
}

@article{shamaei2024water,
  title={Water removal in MR spectroscopic imaging with Casorati singular value decomposition},
  author={Shamaei, Amirmohammad and Starcukova, Jana and Rizzo, Rudy and Starcuk Jr, Zenon},
  journal={Magnetic resonance in medicine},
  volume={91},
  number={4},
  pages={1694--1706},
  year={2024},
  publisher={Wiley Online Library}
}

@article{nagaraja2018tensor,
  title={Tensor-Based Method for Residual Water Suppression in 1H Magnetic Resonance Spectroscopic Imaging},
  author={Nagaraja, Bharath Halandur and Debals, Otto and Sima, Diana M and Himmelreich, Uwe and De Lathauwer, Lieven and Van Huffel, Sabine},
  journal={IEEE Transactions on Biomedical Engineering},
  volume={66},
  number={2},
  pages={584--594},
  year={2018},
  publisher={IEEE}
}

\end{document}